\documentstyle[12pt,aasms4]{article}
\lefthead{JONES, et al.}
\righthead{COUNTERJET IN CENTAURUS A}  
\begin{document}
\title{Discovery of a Sub-Parsec Radio Counterjet in the \\
Nucleus of Centaurus A}  

\author{Dayton L.~Jones\altaffilmark{1}, 
Steven J.~Tingay\altaffilmark{2}, 
David W.~Murphy\altaffilmark{1},
David L.~Meier\altaffilmark{1},\\ 
David L.~Jauncey\altaffilmark{3}, 
John E.~Reynolds\altaffilmark{3},
Anastasios K.~Tzioumis\altaffilmark{3}, 
Robert A.~Preston\altaffilmark{1},\\  
Peter M.~McCulloch\altaffilmark{4},
Marco E.~Costa\altaffilmark{4}, 
Athol J.~Kemball\altaffilmark{5}, 
George D.~Nicolson\altaffilmark{5},\\
Jonathan F.~H.~Quick\altaffilmark{5}, 
Edward A.~King\altaffilmark{4}, 
James E.~J.~Lovell\altaffilmark{4}, 
R.~W.~Clay\altaffilmark{6},\\ 
Richard H.~Ferris\altaffilmark{3},
R.~G.~Gough\altaffilmark{3}, 
M.~W.~Sinclair\altaffilmark{3}, 
S.~P.~Ellingsen\altaffilmark{4}, 
P.~G.~Edwards\altaffilmark{7},\\ 
P.~A.~Jones\altaffilmark{8},
T.~D.~van Ommen\altaffilmark{9}, 
Paul Harbison\altaffilmark{10},
and Victor Migenes\altaffilmark{11}}

\altaffiltext{1}{Jet Propulsion Laboratory, Mail Code 238-332,
California Institute of Technology, 4800 Oak Grove Drive, Pasadena, 
CA 91109, USA; dj@bllac.jpl.nasa.gov}

\altaffiltext{2}{Mount Stromlo and Siding Spring Observatories, 
Australian National University, Private Bag, Weston Creek Post Office,
ACT 2611, Australia; tingay@merlin.anu.edu.au} 

\altaffiltext{3}{Australia Telescope National Facility, CSIRO, 
P.O.~Box 76, Epping, NSW 2121, Australia} 

\altaffiltext{4}{University of Tasmania, G.P.O.~252C, Hobart, Tasmania
7001, Australia} 

\altaffiltext{5}{Hartebeesthoek Radio Astronomy Observatory, 
P.O.~Box 443, Krugersdorp 1740, Transvaal, 
South Africa}

\altaffiltext{6}{University of Adelaide, Adelaide, SA, Australia}

\altaffiltext{7}{Institute of Space and Astronautical Science, 3-1-1,
Yoshinodai, Sagamihara-shi, Kanagawa 229, Japan}

\altaffiltext{8}{University of Western Sydney, Kingswood, NSW, Australia}

\altaffiltext{9}{Antarctic CRC, University of Tasmania, 
Hobart, TAS, Australia} 

\altaffiltext{10}{British Aerospace Australia, Canberra,
ACT, Australia} 

\altaffiltext{11}{National Astronomical Observatory, Mitaka-shi,
Tokyo 181, Japan} 
 
\begin{abstract}  
A sub-parsec scale radio counterjet has been detected in the
nucleus of the closest radio galaxy, Centaurus A (NGC 5128), with VLBI 
imaging at 2.3 and 8.4 GHz.  This is one of the first detections of
a VLBI counterjet, and provides new constraints on the kinematics 
of the radio jets emerging from the nucleus of Centaurus A.   
A bright, compact core is seen at 8.4 GHz, 
along with a jet extending 
along position angle $51^{\circ}$.  The core is  
completely absorbed at 2.3 GHz.  
Our images show a much wider gap between the base of
the main jet and the counterjet at 2.3 GHz than at 8.4 GHz and 
also that the core has an extraordinarily inverted spectrum. 
These observations provide evidence that the innermost 0.4-0.8 pc of
the source is seen through a disk or torus of ionized gas which is
opaque at low frequencies due to free-free absorption.
\end{abstract}

\keywords{accretion, accretion disks --- galaxies: active --- 
galaxies: individual (NGC~5128, Centaurus A) --- galaxies: jets --- 
galaxies: nuclei}

\section{Introduction}

Observations of radio galaxies generally reveal a strong asymmetry in
surface brightness close to the central radio source (associated with
the nucleus of the host galaxy) and more symmetric structure at larger
distances.  An attractive explanation for these observations is that  
the inner structure is intrinsically symmetric -- two 
oppositely directed, relativistic jets of radio emitting plasma -- but
appears asymmetric due to relativistic beaming (amplification) of the emission in
the direction of jet motion.  If the jet material is decelerated
between the nucleus and the large-scale radio lobes, the effect of
beaming is reduced, and the true symmetry of the source becomes 
apparent far from the nucleus.  

A simple expression describes the ratio of observed surface brightness
from approaching and receding radio components 
($R$) in terms of the bulk velocity of material in
the jet ($\beta \equiv v/c$) and the angle between the approaching  
jet axis and our line of sight ($\theta$):
$$R = {\biggl[{{1 + \beta\ {\rm cos}\ \theta} \over
{1 - \beta\ {\rm cos}\ \theta}}\biggr]}^{3-\alpha}$$

\noindent for isotropic emission in the rest frame of the jet 
(Ryle and Longair \markcite{1} 1967).  Here $\alpha$ is the radio 
spectral index of the jet defined as $S_{\nu} \propto \nu^{\alpha}$
where $S_{\nu}$ is the flux density at frequency $\nu$.   
(An exponent of $2-\alpha$ is appropriate if the jet is
modeled as a continuous flow of material instead of a series of
separate knots of radio-emitting plasma [Scheuer and Readhead 
\markcite{2} 1979].)  
Detection of a counterjet places constraints on $\beta$ and
$\theta$, and is important in testing unification schemes based 
on beaming and orientation arguments.  

NGC 5128 is the closest ``classical" radio galaxy, only 3.5 Mpc 
distant (Hui, et al. \markcite{3} 1993). The proximity of this 
active galaxy allows higher linear
resolution imaging than can be obtained for more distant galaxies. 
It is not an 
intrinsically high luminosity radio source, but can be studied over 
a very wide range of linear scales due to its proximity.  The  
extended radio emission associated with Centaurus A covers an area 
of 8.5 by 3.5 degrees on the sky (Junkes, et al. \markcite{4} 1993), 
while observations with 
VLBI arrays can obtain images of the nuclear radio source 
in this galaxy with angular resolution of 1 milliarcsecond or better 
(corresponding to a linear resolution of less than 0.02 pc).
The large-scale radio morphology of Centaurus A, along with its
low intrinsic radio power, suggest a moderate speed jet oriented
close to the plane of the sky.  Therefore this source is a good
candidate for detection of a VLBI counterjet. 

\section{Observations and Results}

At a declination of $-43^{\circ}$, Centaurus A is well placed for
observations with Southern Hemisphere VLBI Experiment (SHEVE) antennas
in Australia and South Africa combined with several antennas of the
Very Long Baseline Array (VLBA).  
This combination provides better coverage of the aperture plane
than either SHEVE or the VLBA alone.  
Our 8.4 GHz observations were made on 20 October 1993 using
the SHEVE array and five VLBA antennas in the southwestern US.  The
resulting image is shown in figure 1.  

\placefigure{fig1}

Three things are immediately noticeable in the 8.4 GHz image:  
the jet is nearly 
linear and oriented along the position angle of the larger-scale jet
seen in VLA images (51$^{\circ}$); 
the peak brightness occurs at the ``base" (southwest end) of 
the continuous jet;  
and there are two isolated peaks of emission approximately 
12 and 27 milliarcseconds southwest of the brightest peak.  We 
associate these peaks with an underlying counterjet because of their 
location along the same position angle as the main jet, on the
opposite side of the core (determined from a comparison of 
4.8 and 8.4 GHz VLBI images, as 
described below).  We also note the presence (in VLA images) 
of other isolated knots of
emission farther from the core and in the direction of the southwest 
inner radio lobe (Clarke, Burns, and Norman \markcite{5} 1992).  

How do we know that one of the features we associate with a
counterjet is not really the core?  
Absolute registration of VLBI images at different frequencies 
is not generally possible,
but a comparison of nearly simultaneous images (separated by 3 days 
during November 1992) at 4.8 and 8.4 GHz shows a distinctive 
``kink" in the main jet 16 milliarcseconds NE of the brightest peak
at 8.4 GHz (see figure 1 in Jauncey, et al. \markcite{6} 1995).  
Aligning this feature
on both images shows that the bright peak at the base of the main
jet at 8.4 GHz has a highly inverted spectrum, while the 
rest of the main jet has a spectrum ranging from less inverted
to flat to steep with increasing distance from the core (Jauncey,
et al. \markcite{6} 1995). 
Other alignments produce nonphysical spectral indices.  
It is possible that the counterjet
also has an inverted spectrum (due, for example, to free-free
absorption in a nuclear disk of gas), but a higher quality 
4.8 GHz image will be needed to test this hypothesis.  The present
4.8 GHz data have a limited dynamic range and are only 
sufficient to constrain the brightest 
regions of the source.  However,  
single antenna flux density measurements show that the brightness
of the nucleus increases rapidly with frequency up to $\approx 22$ 
GHz, indicating that the core does have an inverted spectrum at 
lower frequencies (Beall, et al. \markcite{7} 1978).  Since the 
inverted-spectrum core still 
dominates the total nuclear flux density at 8.4 GHz, 
the brightest peak on our 8.4 GHz image, which is also the only
unresolved feature in the image, is indeed the core. 

Figure 2 shows a SHEVE image from observations made at 2.3 GHz 
on 10 November 1988.  Single-antenna flux density measurements
indicate that the true core is completely absorbed at this
frequency.  The bright jet-like feature in figure 2 can be
identified with the main jet seen at 8.4 GHz, with the brightest
peak at 2.3 GHz identified with the ``kink" in the jet at 
8.4 GHz.  The registration implied by this identification is 
the only physically reasonable one; shifting the 2.3 GHz image by
a single beamwidth produces spectral indices which are implausibly
steep or inverted.  Additional evidence for the registration 
suggested here comes from simultaneous 2.3 and 8.4 GHz VLBA
observations made in July 1995.  These observations are not of
sufficient quality to show the counterjet clearly, but the same
over-all structure as in figures 1 and 2 is evident.  Therefore
morphological changes between the epochs of figures 1 and 2 are
not likely to have affected the large-scale spectral index 
distribution.  

\placefigure{fig2}

\section{Discussion}

It is well known that self-calibration of VLBI data can produce
symmetric structure about bright features even if the true source
brightness distribution is asymmetric (Linfield \markcite{8} 1986).  
This can occur when phases
are self-calibrated with a point source model and either the quantity
of data or the number of self-calibration iterations is insufficient
to correct the initial symmetric bias introduced into the source model.
As a result, most features which appear to be short counterjets in VLBI
images are dismissed as being caused by residual errors in the
self-calibrated data.  The counterjet seen in figures 1 and 2
differs from most calibration artifacts in that it is not  
an extension of (or immediately next to) a bright feature.
 
To test the reality of the counterjet, we
re-imaged the source without allowing any CLEAN components on the
counterjet side during any of the self-calibration and
deconvolution steps.  After the strong
main jet features were included in the model
the brightest residuals were always in the positions of the
counterjet features during every iteration at both 2.3 and 
8.4 GHz.  Allowing CLEAN components
in this region increased the peak brightness of these features by a factor
of two, and improved the over-all fit of the model to the visibilities
by more than 10\% compared to the best one-sided source models.
The counterjet peaks are seen separately
in Mark-II and Mark-III data recorded during the 8.4-GHz experiment.
We have also carried out blind imaging tests using simulated data to
verify that our VLBI array provides sufficient coverage of the aperture
plane to allow reliable imaging of features
similar to those seen in figure 1.  In no case during these blind
tests were counterjets created where none existed in the
starting models. 
Finally, our data were imaged independently by three of the authors,
resulting in virtually identical images.

There is some uncertainty in the registration of the images
in figures 1 and 2,
but any realistic registration shows that the separation between
the core and the first detectable feature in the counterjet is
far greater at 2.3 GHz than at 8.4 GHz.
This is not caused by changes in source morphology during the interval 
between our 2.3 and 8.4 GHz observations unless proper motions within
the counterjet are far greater than those measured for bright features
in the main jet.  
In addition, the spectrum of 
the core is highly inverted ($\alpha \sim 4$)
between these two frequencies.  An analysis of 4.8 and 8.4 GHz images
by Tingay (\markcite{9} 1996) also finds a spectral index $\approx 4$ 
for the core.  Both of these effects can be
explained if the central 0.4-0.8 pc of Centaurus A is seen
through a disk or torus of thermal gas inclined at a large
angle to our line of sight.  A similar situation has been found
in 3C84 (Vermeulen, Readhead, and Backer \markcite{10} 1994; Walker,
Romney, and Benson \markcite{11} 1994).  For example, a 1-2 pc path 
through $10^4$K gas with electron
density of $10^5$ cm$^{-3}$ will give a spectral turnover frequency
$>15$ GHz due to free-free absorption.  The observed spectral index
of about 4 at frequencies below 8.4 GHz would then correspond to an 
intrinsic spectral index of 2.0-2.5 for the core, consistent with 
synchrotron self-absorption.  If this model is correct, the inner 
part of the counterjet  will be reduced in brightness by
a factor of 13 more at 2.3 GHz than at 8.4 GHz, which would explain 
why we do not detect any 2.3 GHz emission at the locations of the
8.4 GHz counterjet features.  

The jet/counterjet brightness ratio $R$ is difficult to determine
from only the brightest peaks of the counterjet.  Comparison of the 
peak counterjet brightness to the main jet brightness at the same 
distance from the core gives $R \simeq 4$ at 8.4 GHz. 
However, if we use the brightest peak along the main jet for our
comparison, values of $R$ up to $\sim 8$ are obtained.  The most
appropriate comparison is between the brightness of the underlying
(smooth?) jet and counterjet emission, not between the brightness
of specific knots, but this must await the availability of higher 
dynamic range VLBI images.  Directly measured proper motions in the
main jet are slow ($\approx 0.15$ c; Jauncey, et al. \markcite{6} 
1995; Tingay, et al. \markcite{12} 1994), but rapid changes   
in the size and brightness of individual jet components suggests 
that the flow speed of the jet is significantly faster 
(${\beta_{\rm app}} \ge 0.85$; Preston, et al. \markcite{13} 1996).
The slower speed is inconsistent with the observed jet/counterjet
bright ratio ($R \le 8$), but the faster speed implied by component
variability is consistent with the brightness 
ratio for a range of angles to our line of sight (60-77$^{\circ}$). 
Such an orientation is consistent with the large-scale radio
morphology of Centaurus A.  This model implies that the bulk flow speed
of the jet need be only moderately relativistic ($\ge 0.66$ c).  

\section{Conclusions}

VLBI observations of many sources have been made to search for 
parsec-scale counterjets, which are predicted by the standard
twin-beam model of radio sources in active galactic nuclei.  However,
the vast majority of radio sources observed with VLBI are high
luminosity (FR II; Fanaroff and Riley \markcite{14} 1974) objects, 
in which the jets are expected to be
highly relativistic and oriented closer to our line of sight (in 
flux limited samples).  Consequently the effects of beaming lead 
to very large values of $R$ and (currently) undetectable counterjets.
Centaurus A, in contrast, is a lower luminosity FR I radio source.  If 
there is a correlation between jet speed and total source luminosity, as
suspected (Bridle and Perley \markcite{15} 1984), then FR I sources 
should have much smaller values of
$R$ than FR II sources at the same angle to our line of sight.
In the case of Centaurus A we also benefit from having a relatively 
large angle between the radio axis (as estimated from the 
large-scale morphology of the source) and our line of sight.  Thus, 
the discovery of a parsec-scale radio counterjet in this source
(and not in many other well-observed sources) is readily explained.  

We are continuing to observe the nucleus of Centaurus A with SHEVE and 
the VLBA.  With multiple epochs, it should be  
possible to measure the apparent proper motion of features in the
counterjet as well as the main jet.  This will provide a direct test 
of symmetric jet models, and further constrain relativistic beaming 
explanations of parsec-scale radio source asymmetries. 

\acknowledgements

We thank the staffs of the observatories participating 
in SHEVE, and the staff of the JPL/Caltech VLBI processor,
for their support of these experiments.  We also thank T.~Pearson
and M.~Shepherd for their development and distribution of the
Caltech VLBI analysis programs.
Part of this research was carried out at the Jet Propulsion
Laboratory, California Institute of Technology, under contract with
the U.S.~National Aeronautics and Space Administration.
The Australia Telescope National Facility is operated in association
with the Division of Radiophysics by CSIRO.

\clearpage

\figcaption{ 
8.4 GHz VLBI image of the nucleus of Centaurus A in 
October 1993, made with antennas in Australia, South
Africa, and the United States.  North is up and east 
is to the left.  The brightest peak corresponds
to the inverted-spectrum core and the counterjet peaks are
located to the southwest of the brightest peak.  
Note also the slight wiggle or kink in the main jet 
15-20 mas northeast of the brightest peak.  
The contour levels are -1, 1, 2, 4, 8, 16, 32, 
50, 70, and 95\% of the peak brightness (1.80 Jy/beam).  
The restoring beam (shown in the lower left corner) 
is $3.22 \times 1.82$ 
milliarcseconds with the major axis in position
angle $26.2^{\circ}$. \label{fig1}} 

\null

\figcaption{ 
2.3 GHz SHEVE image of the nucleus of Centaurus A in November  
1988, made with six antennas in Australia.  The contour levels 
are -1, 1, 2, 4, 8, 16, 32, 50, 70, 
and 95\% of the peak brightness (1.43 Jy/beam).  The orientation
and scale of this figure are the same as in figure 1.  The restoring beam 
is $7.26 \times 5.53$ milliarcseconds with the major axis in position
angle $8.5^{\circ}$.  The brightest peak in this image corresponds
to the ``kink" in figure 1, and the approximate location of the 
(absorbed) core is marked. \label{fig2}}  

\end{document}